\documentclass[manuscript]{aastex}

\received{}
\revised{}
\accepted{}

\shortauthors{Jones}
\shorttitle{Magnetic Fields in M82 and Cen A}

\begin{document}

\title{The Magnetic Field Geometry in M82 and Cen A}
\author{Terry Jay Jones\altaffilmark{1}}

\affil{Department of Astronomy, University of Minnesota}
\affil{Minneapolis, MN 55455} 
\email{tjj@astro.spa.umn.edu}

\altaffiltext{1}{Visiting Astronomer at the Infrared Telescope
Facility
which is operated by the University of Hawaii under contract from the
National Aeronautics and Space Administration. }

\begin{abstract}
Imaging polarimetry at 1.65 and 2.2 $\mu $m is presented for the
classic starburst galaxy M82 and the advanced merger system Cen A.
Polarimetry at near IR wavelengths allows the magnetic field geometry
in galaxies to be probed much deeper into dusty regions than optical
polarimetry. In M82, the magnetic field throughout the nucleus has a
polar geometry, presumably due to the massive vertical flow that
is a result of the intense star formation there. Fully two thirds of
the line of sight dust through to the center of M82 contains a
vertical magnetic field. In Cen A, the prominent dust lane shows a
normal planar field geometry. There is no indication of significant
disturbance in the field geometry in the dust lane and the
polarization strength is near normal for the amount of extinction.
Either the magnetic field geometry was well maintained during the
merger, or it reestablished itself very easily.
\end{abstract}


\section{Introduction}

Magnetic fields play an important role in the star formation process
\citep{cru99} and in the motions of gas within galaxies
\citep{val97}. This must certainly be the case for the elevated
activity and violent gas motions taking place in starburst nuclei.
Magnetic fields must also interact with gas motions in disturbed disk
systems and mergers, but their role in merging systems is less clear. By
measuring the magnetic field geometry in starburst nuceli and disturbed
disk systems, we can find important clues to large scale gas motions that
have recently taken place there. However, the geometry of the magnetic
field in these environments is difficult to determine as is not well
known.

Faraday studies at radio wavelengths can not effectively probe the
dense gas seen along the line of sight to active nuclei, particularly
for edge-on systems. Polarimetry at optical and infrared wavelengths
can provide direct information on the magnetic field geometry in most
phases of the ISM with the probable exception of most dense, cold
dark clouds \citep{goo96, jon96}. Consequently polarimetry of light
from stars in external galaxies shining through the {\it diffuse} ISM can
be an excellent probe of the integrated projected magnetic field
geometry in those systems.

Polarimetry at infrared wavelengths has the advantage that it
penetrates deeper into dusty regions of spiral galaxies and heavily
obscured nuclear regions than is possible at visual and UV
wavelengths. Infrared imaging polarimetry of the normal edge-on
spiral NGC 4565 by \citet{jon97} clearly shows a simple toroidal
magnetic field geometry lying in the plane of the galaxy, similar to
the field in the Milky Way. \citet{woo97} were able to model these
observations incorporating scattering and polarization in
transmission using the model developed by \citet[JKD]{JKD} from Milky
Way observations. These studies confirm that infrared imaging
polarimetry of relatively quiescent normal spiral galaxies does map
out the magnetic field geometry in the disk and is in good agreement
with the polarization characteristics seen in the Milky Way.

In this paper we present infrared imaging polarimetry of the classic
starburst galaxy M82 and the late merger system Cen A. We have chosen
M82 to investigate the effects of a vigorous starburst on the
magnetic field geometry. Cen A was chosen to probe the field geometry
in its peculiar dust lane, which is widely thought to be due to a
merger of the elliptical galaxy with a dusty disk system.  We show
that the observations of M82 are consistent with a completely polar
magnetic field geometry in the inner 300 parsecs of the galaxy. For
Cen A we find that the polarization in the prominent dust lane is
nearly normal for a quiescent disk, and shows no sign of significant
memory of a past merger.

\section{Observations}

The observations were made using NSFCAM on the IRTF. The basic technique
for using NSFCAM as an imaging polarimeter is given in \citet{jon97}. For
the new observations reported here, we changed the observing technique
slightly to reduce systematic errors. Instead of measuring an object-sky
pair at one position of the waveplate followed by a pair at the next
position of the waveplate (45$^{\circ }$ of waveplate rotation equals
90$^{\circ }$ in polarization), we stepped back and forth between
waveplate positions on the object for a while, then on the sky. This
change in technique essentially sped up the polarization sampling rate,
reducing our sensitivity to systematic effects due to variations in system
response. For sources with blank sky on the frame, our polarization
precision was improved by about a factor of two. We could reach a
precision of +/- 0.1\% on the best nights. For more extended sources, the
lack of pure sky on a single image frame complicates the data reduction
somewhat since the sky frames are observed several minutes after the
source frames. This made accurate subtraction of the sky brightness (which
varies during the night) more difficult, but we were still able to obtain
an improvement over the previous technique.

The data reduction analysis was, otherwise, similar to that described in
Jones (1997). Instrumental polarization was checked by observing
unpolarized stars and was unmeasurable at the +/- 0.1\% level.
Polarization efficiency and calibration of the position angle was
accomplished by observing the star S1 in $\rho $ Oph which is assumed to
have P = 3.90\%, $\theta $ = 28$ ^{\circ }$ at H band (1.65 $\mu $m).
NSFCAM in polarimetry mode is over 97\% efficient at K (2.2 $\mu $m) and
95 +/- 0.2\% efficient at H. The seeing was measured to be 0.9'' FWHM on
all nights. The intrinsic plate scale for NSFCAM was 0.3'' per pixel, and
we binned the images to larger angular sizes to improve the
signal-to-noise. Near IR $H-K$ colors for the two galaxies were derived
from separate K band images taken through the polarization optics (but
without rotating the waveplate). The H band images used in computing the
colors were simply the sum of the many images taken in polarimetry mode.
Calibration of the colors was done using JHK standards from the UKIRT
faint standards list available at the IRTF. These photometric standards
were observed at H and K in polarimetry mode. Note that NSFCAM is
incapable of imaging polarimetry in the J (1.25$\mu $m)photometric band
because a long wavelength blocker for the J filter is in the same wheel as
the wire grid polarizer. For this reason we have no J band polarimetry.

The observation Log is presented in Table 1. About 1.5 hours of telescope
time was spent on each object on each night it was observed. For
polarizations greater than about 0.4\%, our polarization error is
consistent with the background photon noise. In other words, the
photometric accuracy constrained by the background photon flux is the
determining factor in the polarization precision. If the surface
brightness is low, then the polarization error will, of course, be high
due to lack of signal strength. For polarizations less than about 0.4\%,
systematic effects are an increasing contributing factor, even when the
surface brightness is high and there is plenty of signal. Our final
precision for the map of M82 ranges from +/-0.1\% in the central portions
of the image to about +/-0.2\% in the outer portions of the image where
the surface brightness is lower. For Cen A, the precision is about +/-
0.2\% across the entire image when binned to 2.7'' pixels. The precision
is better for M82 because we spent more time observing that object and
because one of the nights had exceptionally low fluctuations both in sky
transmission and in the sky background (mostly airglow in the H band)
during the time M82 was being observed. For Cen A, fluctuations in the H
band airglow and low surface brightness prevented our precision from
reaching +/- 0.1\%.

\section{M82}

\subsection{Introduction}

M82 is the archetypal starburst galaxy, extensively studied at all
wavelengths \citep{tel88}. It contains a superwind originating from
the nuclear region where the major fraction of the massive starformation
is taking place. This wind produces a relatively unique bipolar
reflection nebula above and below the plane of the edge-on system clearly
seen in imaging polarimetry at visible wavelengths \citep{sca91}. Optical
imaging polarimetry shows a hint of polarization in transmission from the
central portions of the galaxy \citep{nei90}. This polarimetry suggested
the magnetic field geometry is perpendicular to the plane of the galaxy,
unlike normal spirals where the field lies in the rotating disk. At
infrared wavelengths, it is much easier to penetrate the large extinction
towards the nucleus of M82 and sample the interstellar polarization along
the entire path to the nucleus. \citet{die89} presented aperture
polarimetry of the nucleus at 1.65 and 2.2 $\mu $m that clearly showed a
polar field geometry for the magnetic field.

Recently \citet{gre00} presented submm polarimetry of M82 at 850$\mu $m
with a beam size of 15\arcsec. They find a somewhat confused pattern for
the field geometry close to the nucleus, but find a more centrosymmetric
field pattern surrounding the nucleus and the molecular torus. Since submm
polarimetry samples polarized {\it emission}, (and the polarization
vectors for this 'halo' actually point radially out from the nucleus),
this centrosymmetric pattern can not be due to scattering of light from
the very luminous nucleus and torus, but must represent the magnetic field
geometry in the diffuse gas surrounding the nucleus. \citet{gre00} suggest
a magnetic bubble has been blown by the intense activity in the nucleus.
Our near IR polarimetry is easily contaminated by scattered light and will
be unable to corroborate the geometry found by \citet{gre00}, but we are
able to probe the magnetic field geometry directly in the nucleus and well
off the nucleus in the galactic plane at much higher spatial resolution. 

\subsection{Discussion}

A polarization map of M82 at 1.65 $\mu $m is shown in Figure 1 at 1.5
\arcsec spatial resolution. A grey scale intensity image at the same
wavelength underlies the polarization map. Figure 2 is a polarization map
of the central region of the galaxy at 0.9\arcsec spatial resolution (the
seeing limit) with intensity contours underlying the map. A strong
scattering halo is clearly present in the data. Since the extinction is
less at 1.65 $\mu $m than at V ($A_{V}\sim 6A_{H}$), light from the very
luminous nucleus can penetrate into the disk and scatter into our line of
sight from most any location. For this reason the scattering halo makes a
full circle around the nucleus. Although it is entirely possible for some
of the centrosymmetric pattern seen surrounding the nucleus of M82 to be
caused by transmission through grains aligned with the centrosymmetric
field found by \citet{gre00}, it would be virtually impossible to separate
scattering and transmission effects in our maps. Consequently, we will
concentrate on the nucleus itself and the edge-on disk to the South West.

Careful examination of the polarization map reveals polarization in
transmission from the central 10\arcsec of the galaxy and from the
edge-on disk in the SW. This can be seen more clearly if the polarization
vectors dominated by scattering are removed. We can do this by retaining
only those polarization vectors with a position angle departing more than
45$^{\circ }$ from the value expected for pure centrosymmetric scattering.
Regions not dominated by scattering of light from the nucleus will be left
in evidence. Figure 3 illustrates the results of this procedure, which
removed about 60\% of the total number of polarization vectors in the map.
This high percentage of vectors removed further illustrates the extent to
which scattering dominates the polarization of M82. If a threshold smaller
than 45$^{\circ }$ is used, fewer vectors will, of course, be removed.
Significantly larger values for the threshold will remove a few vectors in
the NE and in the region directly South of the nucleus. The polarization
vectors in the nucleus and in the SW disk are basically perpendicular to a
centrosymmetric geometry and can not be due to scattering. Figure 3 is
primarily intended to highlight these polarization vectors, not to provide
a quantitative measure of departure from centrosymmetric scattering.

Clearly evident in Figure 3 is polarization in transmission from the
nucleus that has a position angle nearly perpendicular to the disk of the
galaxy. The position angle is not exactly perpendicular to the disk
however, rather it is offset about 15$^{\circ }$ to the East. A similar
offset is seen in the bipolar nebulosity imaged in emission lines at
visible wavelengths by  \citet{sho98}, which they interpreted as an
asymmetric wind. In other words, our polarization measurements show the
same shift in the angle for the outflow axis as do the optical emission
line images. We conclude that the interstellar polarization tracing the
magnetic field geometry in the nucleus of M82 is delineating a poloidal
field aligned with the bipolar nebula and the superwind emanating from the
starburst nucleus. 

To the SW, the planar field in the disk is clearly evident at a position
angle of about 55$^{\circ }$. Consider the rectangular region on the sky
bounded by (-15, -5) at the NE corner and (-40, -20) at the SW corner. A
histogram of the position angles for the polarization in this region is
shown in Figure 4. The dust lane is evident by the peak in the
distribution centered at 55$^{\circ }$. The central peak in the
distribution at 55$^{\circ }$ has a dispersion (standard deviation) of
about 10$^{\circ }$, very similar to the dispersion seen in the disk of
the normal edge on spiral NGC 4565 \citep{jon97}.

The simple model developed by \citet{jon92} to explain interstellar
polarization in active galaxies is illustrated in Figure 5. Galaxies
with intrinsically polarized AGN are not considered in this model. For
this paper we have added a scattering halo to the model, in hindsight
based on our results for M82. Note that the light from the nucleus
traverses two regions of field geometry. The nuclear region has a field
blown polar by the superwind emanating from the center. The foreground
disk has a normal toroidal geometry for disk systems. This produces a
'crossed polaroid' effect and reduces the net polarization. In this simple
picture, the region with the greatest extinction will determine the net
position angle of polarization. If the extinction through the disk and the
nucleus are nearly equal, the net polarization will be nearly zero. If the
extinction through the nucleus (for example) is greater, the net
polarization will still be less than expected for the total extinction,
but the position angle will be polar. If the extinction through the disk
is greater, the polarization will again be weak but the position angle
will lie in the plane.

Following \citet{jon92} we can estimate the fraction of the total
extinction along the line of sight to the far side of the nucleus that is
resides in a volume with a polar field geometry. Since the net position
angle is polar, the larger fraction of the extinction is in the nuclear
region as opposed to the foreground disk, which presumably has a planar
field geometry (as seen in the SW). The relative proportion of the
extinction between the two regions can be estimated from the amount by
which the polarization is reduced below the expected strength for the
measured extinction. An estimate of extinction is obtained from the $H-K$
color assuming an intrinsic color typical of the inner bulge of a galaxy
($(H-K)_{0}=0.25$, \citet{gri82}). HII emission can contaminate the $H-K$
colors and make them appear redder than they would be due to extinction
alone, but we are assuming starlight dominates at these wavelengths in the
nucleus and disk of M82 (see the discussion in \citet{jon97}). Our $H-K$
color for the nucleus ($ E(H-K)=0.55\rightarrow A_{V}\sim 9$) is
consistent with (or at the low end of) other estimates of extinction that 
are independent of HII emission \citep{tel88}.

Our previous aperture polarimetry \citep{jon92, die89} suggested that
fully 2/3 of the extinction path along the line of sight through the
foreground disk to the far side of the nucleus of M82 contained a polar
field geometry and this is confirmed by our imaging polarimetry. From our
imaging polarimetry, we can confidently state that the inner 250-300
parsecs of M82 contains a polar field geometry. The integrated column of
dust through the region with a polar field geometry is fully 2/3 of the
total dust column through to the far side of the nucleus of M82. The
remaining 1/3 is presumably in the foreground disk (the disk behind the
nucleus is probably too faint to contribute much light). In other words,
virtually ALL of the gas in the nuclear region of M82 has a poloidal field
geometry. This strongly suggests that most of the roughly 10$^{8}M_{\odot
}$ of gas (see discussion in \citet{mcl93}) in the nuclear region is
moving perpendicular to the plane. This phenomenon is taking place over a
large volume at least 250 pc in diameter, not just from a very compact
source located at the center (see also \cite{alt99}).

\section{Cen A}

\subsection{Introduction}

Cen A is the closest powerful radio galaxy. It is a well known radio
variable with bipolar lobes, a jet, optical filaments and a prominent
dust lane across the bright elliptical galaxy. A comprehensive review of
Cen A can be found in \citet{isr98}. Most recent studies of Cen A have
concentrated on the nature of the AGN at the center and its effects on its
immediate environment. In this study, we will investigate the magnetic
field geometry in the large dust lane that stretches across the face
of the elliptical galaxy.

Near infrared aperture polarimetry of the nucleus of Cen A was analyzed by 
\cite{bai86}. The normal interstellar polarization caused by the
foreground dust lane was a polarization component that needed to be
removed in their study of the AGN. They found that the position angle of
the interstellar component was coincident with the direction of the dust
lane. This is the expected geometry for a galactic disk. More recently,
imaging near infrared polarimetry by \citet{pac96, pac98} extended
Bailey's work. The polarization due to transmission through the dust lane
is clearly evident in their images and is largely aligned with the
direction of the dust lane. Note that near IR polarimetry of a supernova
in Cen A \citep{hou87} showed the same wavelength dependence for
interstellar polarization as found in the Milky Way. This strongly
suggests direct comparisons between the interstellar polarization in Cen A
and the Milky Way are justified.

\subsection{Discussion}

A polarization map of the inner 1' of Cen A at 2.7\arcsec spatial
resolution at 1.65 $\mu $m is shown in Figure 6. A grey scale image of the
galaxy at the same wavelength underlies the polarization map. No
polarization vectors with a S/N of less than 3/1 are plotted. A
polarization map of the inner 30 \arcsec(where the signal strength is
highest) is shown in Figure 7 where the spatial resolution is 0.9\arcsec.
Intensity contours underlay the map.

Note the polarization along the dust lane in Figure 6 and the twist in the
position angle of the polarization at the nucleus in Figure 7. This twist
is due to the combined effect of the intrinsic polarization for the
nucleus and the interstellar polarization in the intervening dust lane
being at different angles. This observation is extensively discussed in
\citet{pac96,  pac98}. We are interested in the interstellar polarization
in the dark dust lane that stretches from the NW down to the SE, passing
about 8 \arcsec South of the nucleus. Note that in optical images the dust
lane appears much wider than in our longer wavelength image. 

We isolate a 10\arcsec wide strip stretching from coordinates (30, 5) to
(-30,-20) for analysis. This strip contains the most opaque region in the
dust lane for Cen A.  The distribution of position angles for the
polarization vectors in this dust lane is shown in Figure 8. The mean of
the distribution (115$^{\circ }$ ) corresponds nicely to the position
angle of the dust lane. The dispersion of the polarization position angles
is $\sigma =9.5^{\circ }$ and this includes extra dispersion due to
measurement error. Taking into account the errors in our polarimetry, the
intrinsic dispersion for the polarization position angles along the dark
dust lane in Cen A is about 7$^{\circ }$, essentially identical to the
intrinsic dispersion found for NGC 4565 \citep{jon97}. In other words, the
magnetic field geometry in the most opaque portion of the dust lane of Cen
A shows no sign of chaotic or systematic dispersion in position angle
beyond that seen in normal spiral galaxies. 

\section{Comparisons}

A comparison of the observed polarization 'efficiency' for several
galaxies with near infrared imaging polarimetry is shown in Figure 9
\citep[this paper]{jon97}. Figure 9 plots the observed polarization vs.
the observed color excess $E(H-K)$, which is a measure of the extinction
into the galaxy. The determination of the color excess is discussed in
\citet{jon97}. The numerous circles and ovals enclose the region of the
plot occupied by the observations of each individual galaxy or a specific
region in that galaxy. To be more precise, the ovals enclose 75\% of the
data points for each system. Plotting all of the data with different
symbols for each galaxy produces a plot that is impossible to decipher.

The two dashed lines are the results of model calculations using the
JKD model developed to describe interstellar polarization in the Milky
Way. The JKD model included a random component to the magnetic field and
was 'calibrated' using the observed interstellar polarization in the
Milky Way. In the Milky Way, individual stars are observed through an
intervening screen of dust. These observations cluster about the upper
model line labeled JKD Screen, which assumes individual lines of sight to
individual stars. Observations of interstellar polarization in the disks
of external galaxies, however, sample many stars and their individual
line-of-sight extinction within a beam. A more appropriate description of
the geometry of the stars and dust in a galaxy disk is a uniform mix of
stars and dust. 

The JKD model can be used to predict the behavior of such a geometry. This
variation on the model essentially computes the effects of observing
starlight emerging from a slab of uniformly mixed stars and dust. The
total optical depth of the slab is varied, and this produces the range in
$E(H-K)$ colors and net polarization strength plotted as the line labeled
'JKD Mix' in Figure 9. Computationaly this is achieved by dividing up the
optical depth into intervals exactly as in the JKD model, and simply
adding a source term at each interval. Note that this model saturates at
$E(H-K)\sim 0.65$ where the optical depth into the mix of dust and stars
at H and K significantly exceeds unity. Of course, any specific
beam on a galaxy may contain some combination of both geometries. A bright
nucleus behind a dust lane, for example, will behave more like a dust
screen than a mix of stars and dust. None-the-less, these two models
probably span the maximum and minimum theoretical polarization assuming a
single direction for the magnetic field geometry.

The two normal edge-on spirals in our sample, NGC 4565 and NGC 5746
are well explained by the JKD Mix model. These normal galaxies must have
magnetic field geometries and fluctuations in the magnetic field that are
very similar to the Milky Way. NGC 891, which has a very active disk with
strong evidence for large scale blowouts (e.g. \cite{how99}), is
significantly weaker in polarization for the amount of extinction in the
disk. \citet{jon97} interpreted this as due to a vertical component to the
magnetic field in the disk associated with the large regions of blowouts
and suprebubbles. The net position angle of the polarization is still
mostly in the plane, indicating the normal toroidal field occupies the
larger fraction of the extinction path into the disk of that galaxy.

The inner portions of M82 are also weakly polarized for the observed
extinction. In this case, the net position angle is perpendicular to
the plane, indicating that the poloidal field in the nucleus occupies the
larger fraction of the extinction path length. The disk of M82 to the SW
also shows weak interstellar polarization but the position angle is in
the plane. This is similar to the case for NGC 891 \citep{jon97} an
edge-on spiral with a very active disk. We interpret the disk in the
SW of M82 in the same way as NGC 891, having regions with significant star
formation that have produced several large scale vertical motions, but
still containing less gas and dust than the surrounding quiescent disk
material. This interpretation must be tempered by the very real
possibility that scattered light from the nucleus may be contaminating the
polarization in the SW plane of M82, and artificially reducing the
observed polarization strength. Scattered light can not, however, produce
the observed position angle in SW disk of M82. This must be due to
polarization in transmission.

Cen A shows a polarization strength only slightly lower than expected
for the extinction. This indicates there is relatively little departure
from a normal toroidal field geometry in the dustlane. The dispersion in
the position angles of the polarization about the direction of the plane
is as low as in NGC 4565, also indicating little disturbance beyond the
normal random component to the interstellar magnetic field seen in the
Milky Way. Although Cen A is commonly considered to be in the late stages
of a merger, the magnetic field geometry in the disk shows no signs of
such a history at the present epoch.

\section{Conclusions}

1) In M82, centrosymmetric scattering of light from the nucleus strongly
dominates the polarization map. Polarization in transmission is evident at
the nucleus and in the disk to the SW.

2) In M82, the magnetic field in the nucleus has a polar geometry,
presumably due to the massive vertical flow that is a result of the
intense star formation there.

3) Virtually all of the gas and dust in the inner 300 pc of M82 contains a
vertical magnetic field.

4) In Cen A, the prominent dust lane shows a normal planar field geometry.
There is no indication of significant disturbance in the field geometry of
the dust lane and the polarization strength is near normal for the amount
of extinction.

\section{acknowledgments}

We would like to thank the support staff of the IRTF for their assistance
helping make imaging polarimetry a viable option on the IRTF.

\newpage

\figcaption[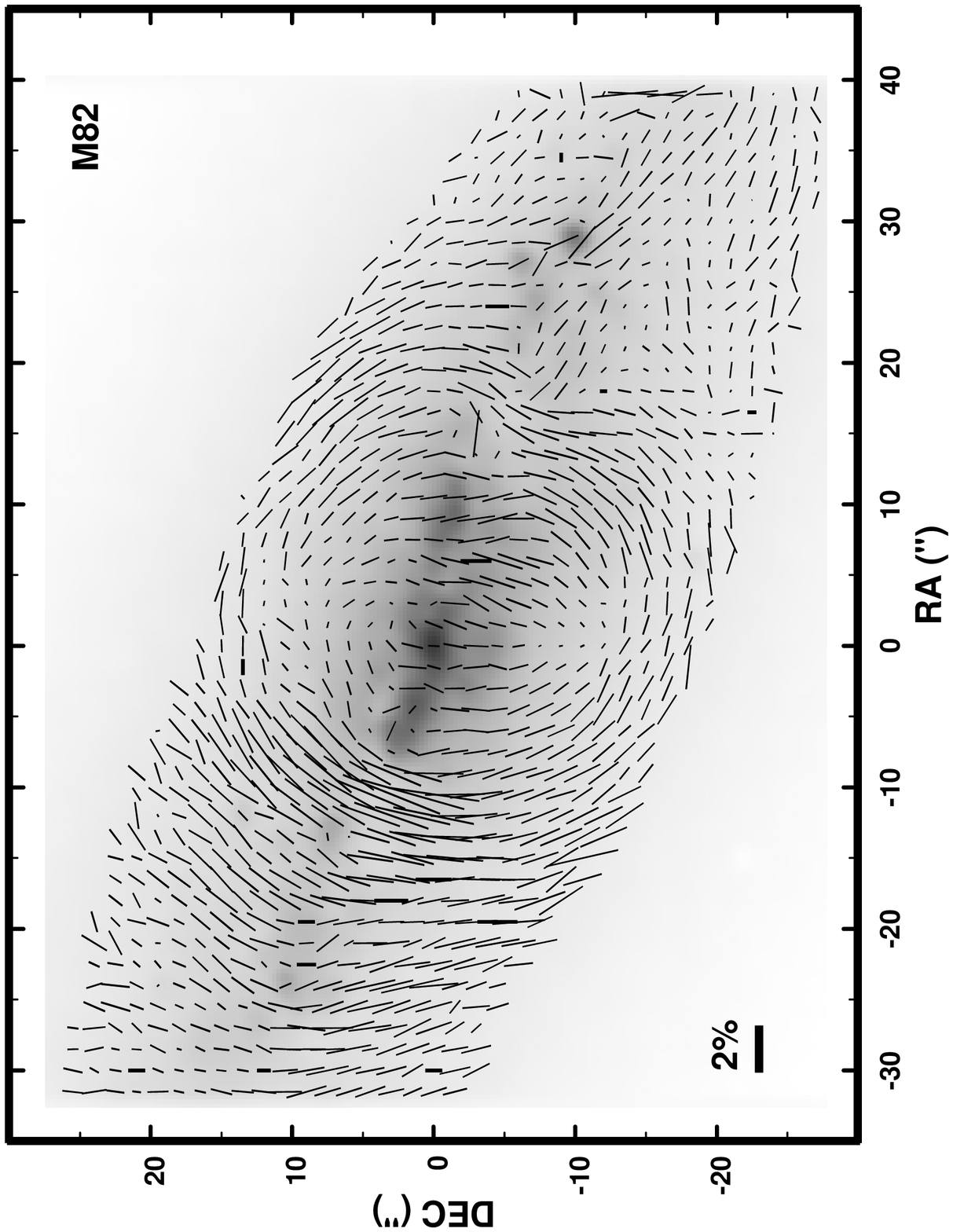]{A polarization map of M82 at H (1.65
$\mu $m) overlaying a grey scale image at the same wavelength. The data
has been binned to a resolution of 1.5\arcsec. \label{fig1}}

\figcaption[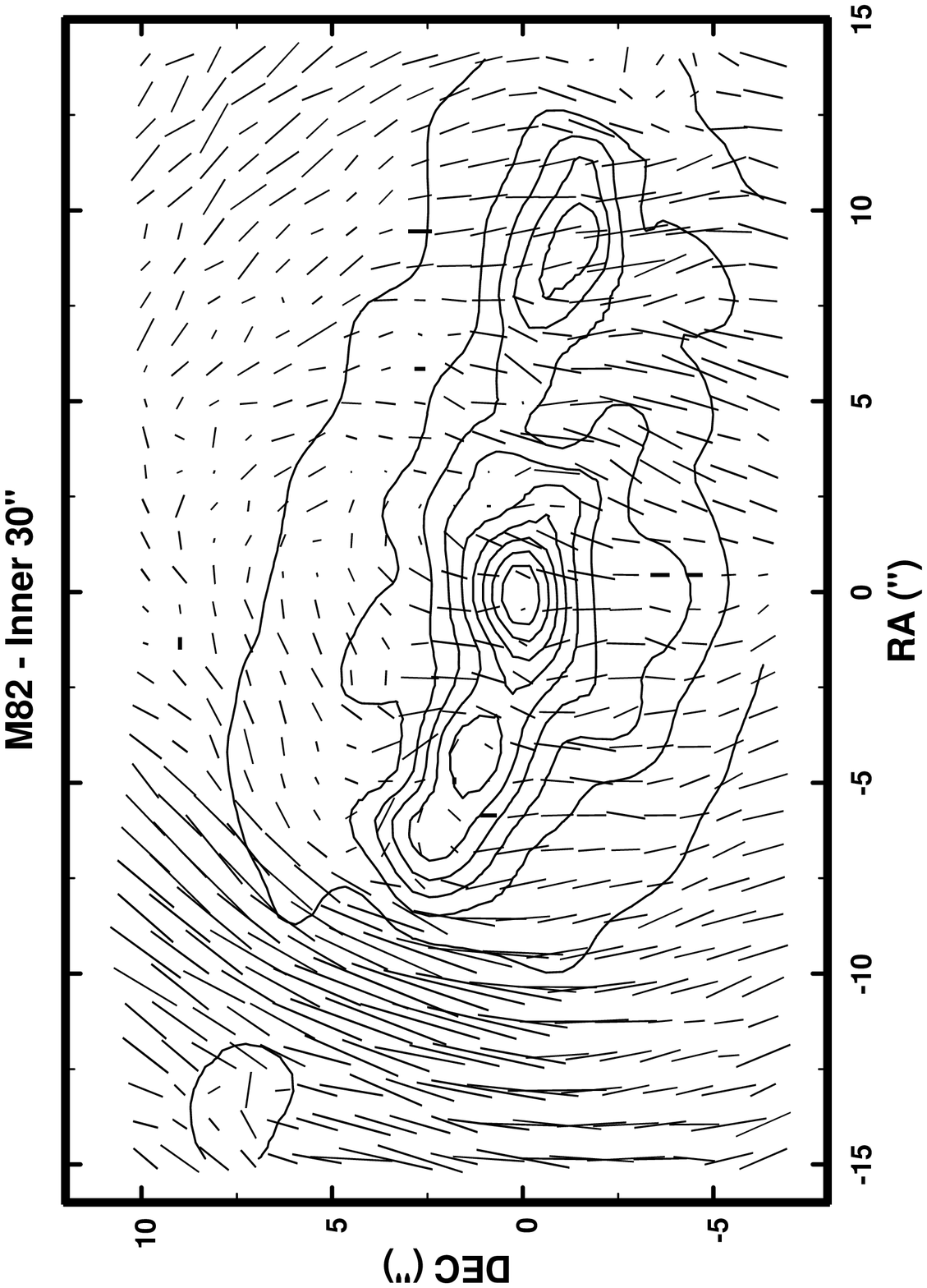]{A polarization map of the inner portions
of M82 at a spatial resolution of 0.9\arcsec, the limit of the seeing. The
underlying contours are intensity levels at H. \label{fig2}}

\figcaption[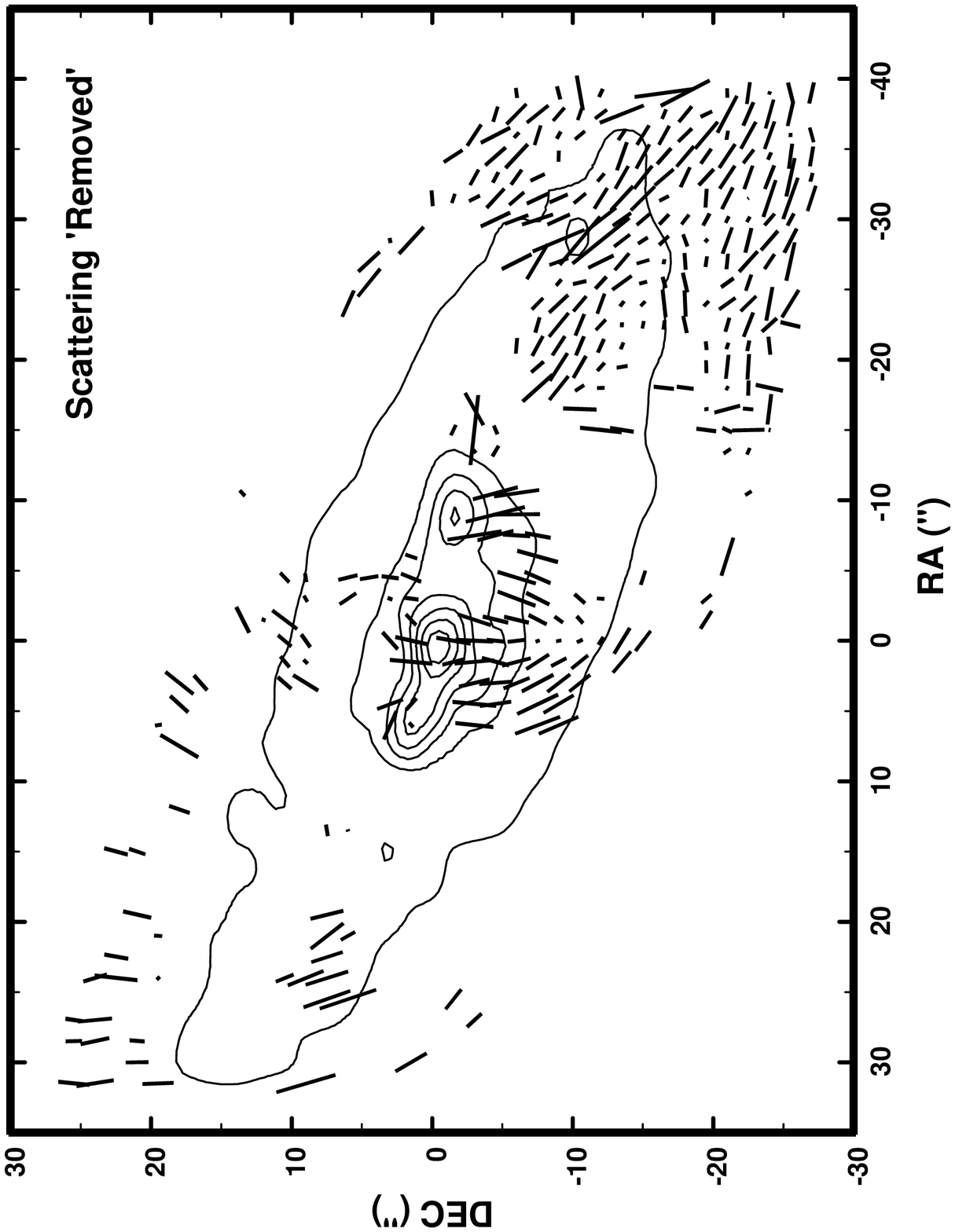]{A polarization map of M82 after removal of
all polarization vectors consistent with centrosymmetric scattering. This
leaves those polarization vectors that can not be exclusively due to
scattering of light from the nucleus. \label{fig3}}

\figcaption[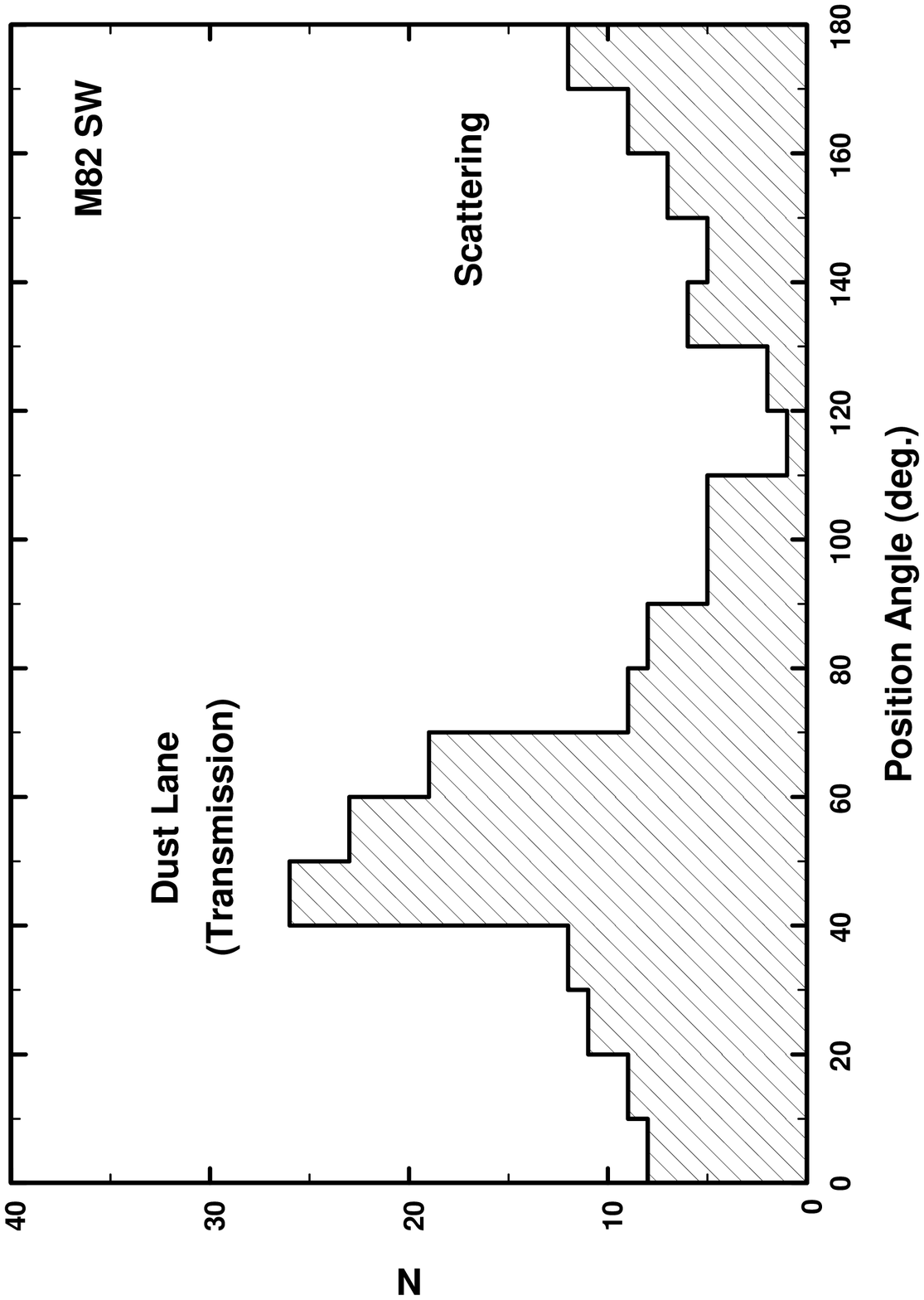]{A histogram of the distribution of the
position angles for the region in the SW of M82 bounded by the coordinates
(-15, -5) and (-40, -20). This boundary selects the region in the SW of
Figure 4 that is dominated by interstellar polarization lying in the disk
of M82. \label{fig4}}

\figcaption[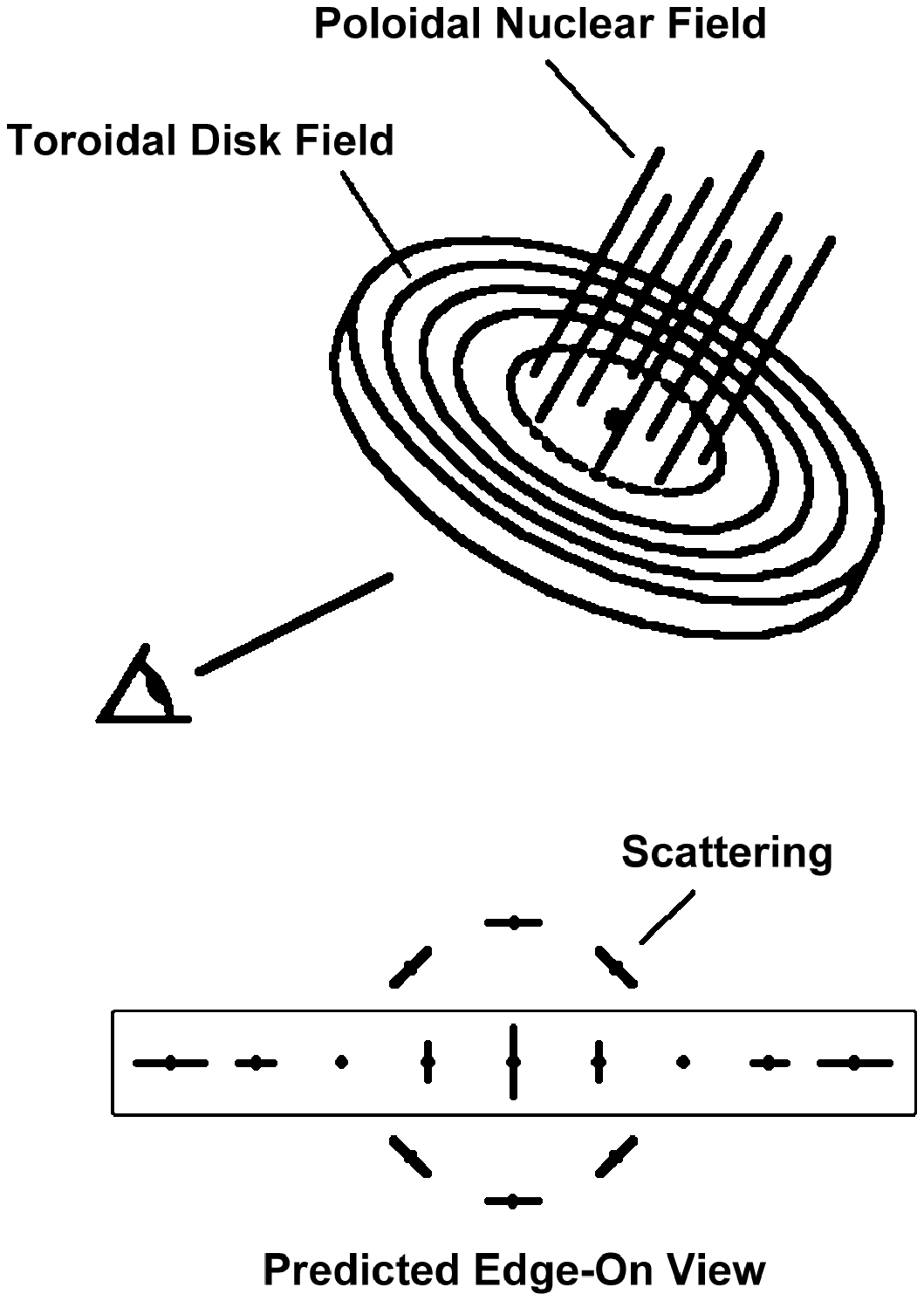]{Simple model to explain the observed
interstellar polarization in active galaxies (from Jones 1992).
\label{fig5}}

\figcaption[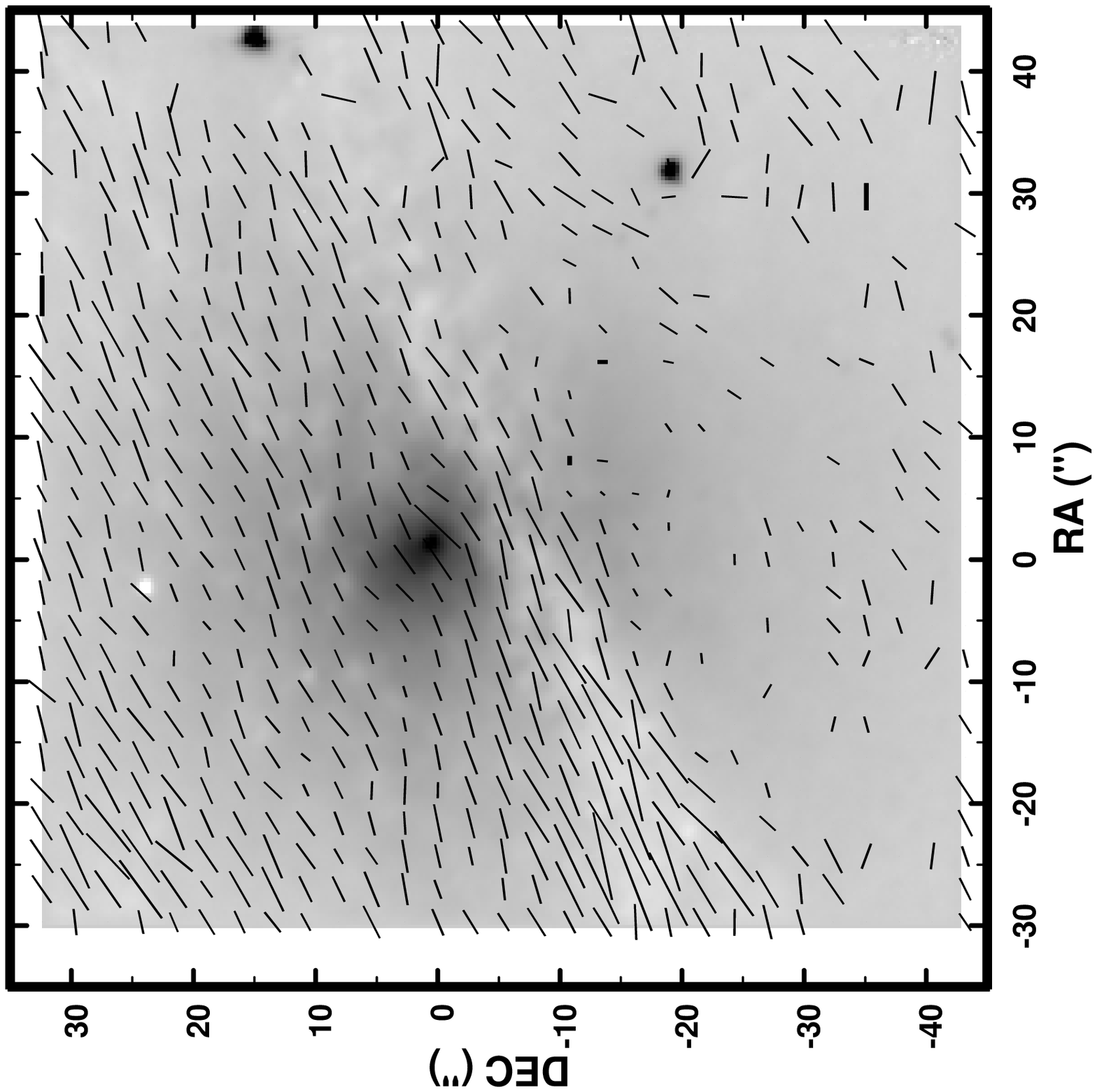]{A polarization map of Cen A at H overlaying a
grey scale image at the same wavelength. The data has been binned to a
spatial resolution of 3\arcsec. \label{fig6}}

\figcaption[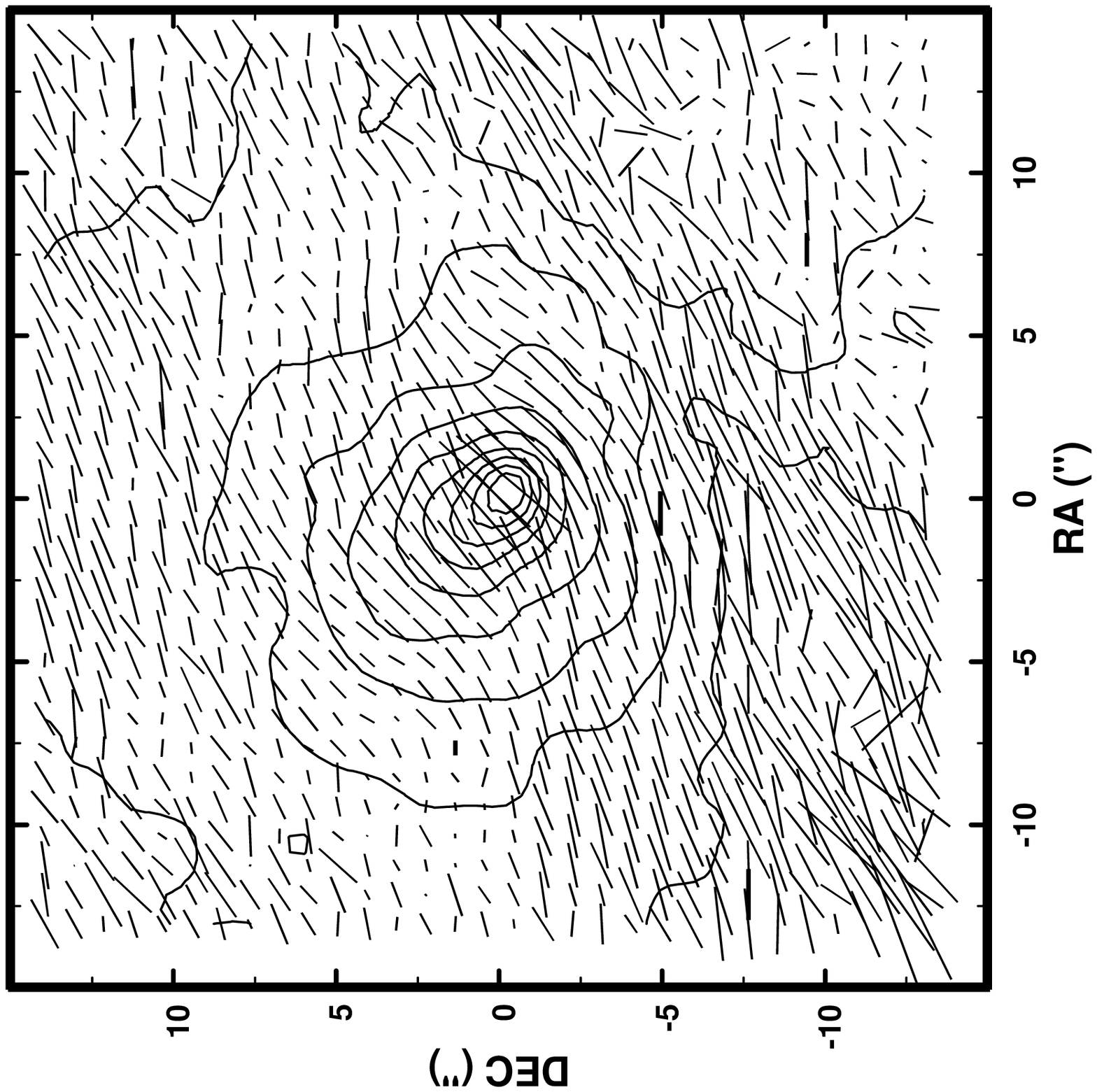]{A polarization map of the inner portions of
Cen A at H at a spatial resolution of 0.9\arcsec, the seeing limit. The
underlying contours are intensity levels at H. \label{fig7}}

\figcaption[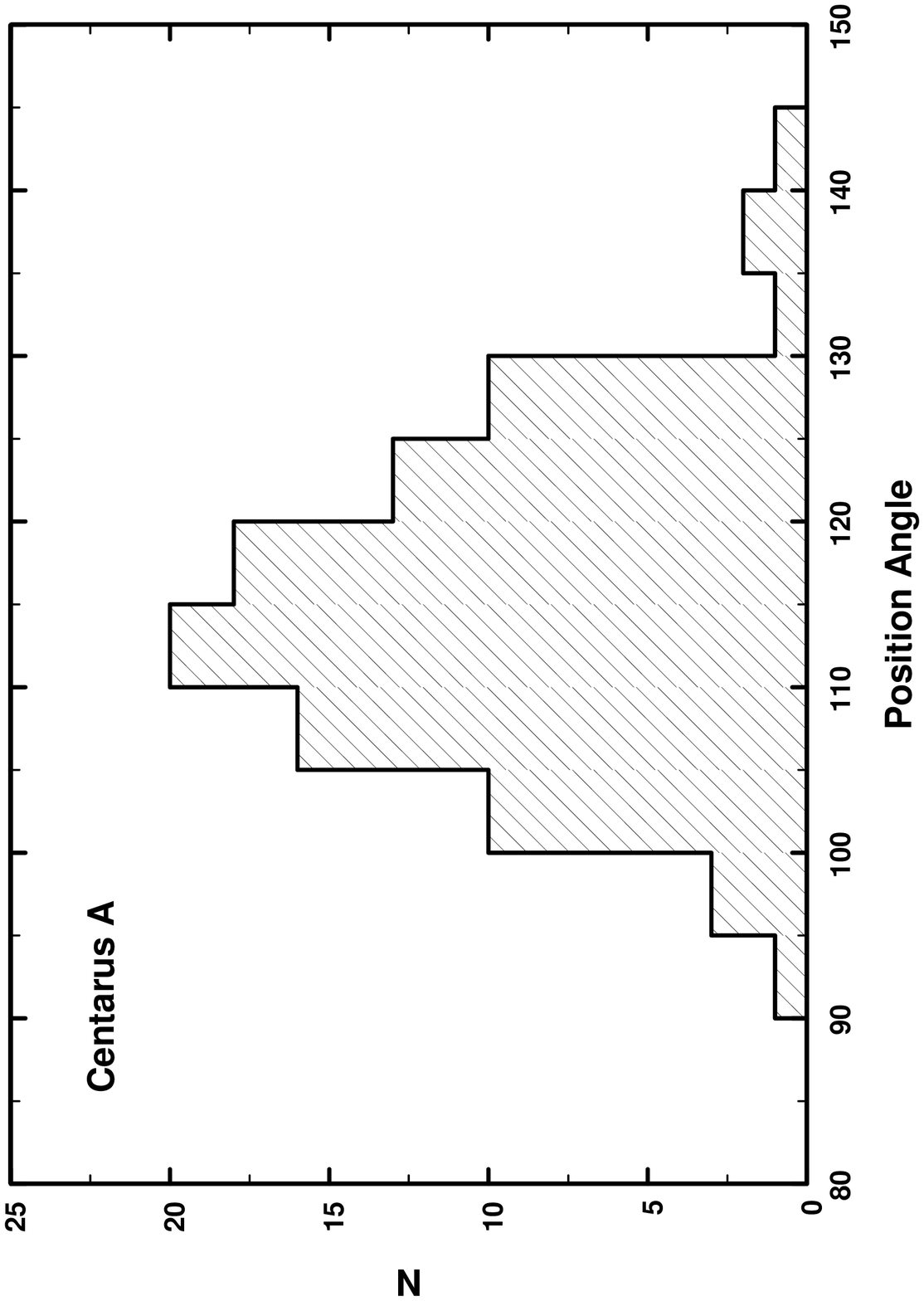]{A histogram of the distribution of the
position angles for the darkest dust lane in Cen A (see text).
\label{fig8}}

\figcaption[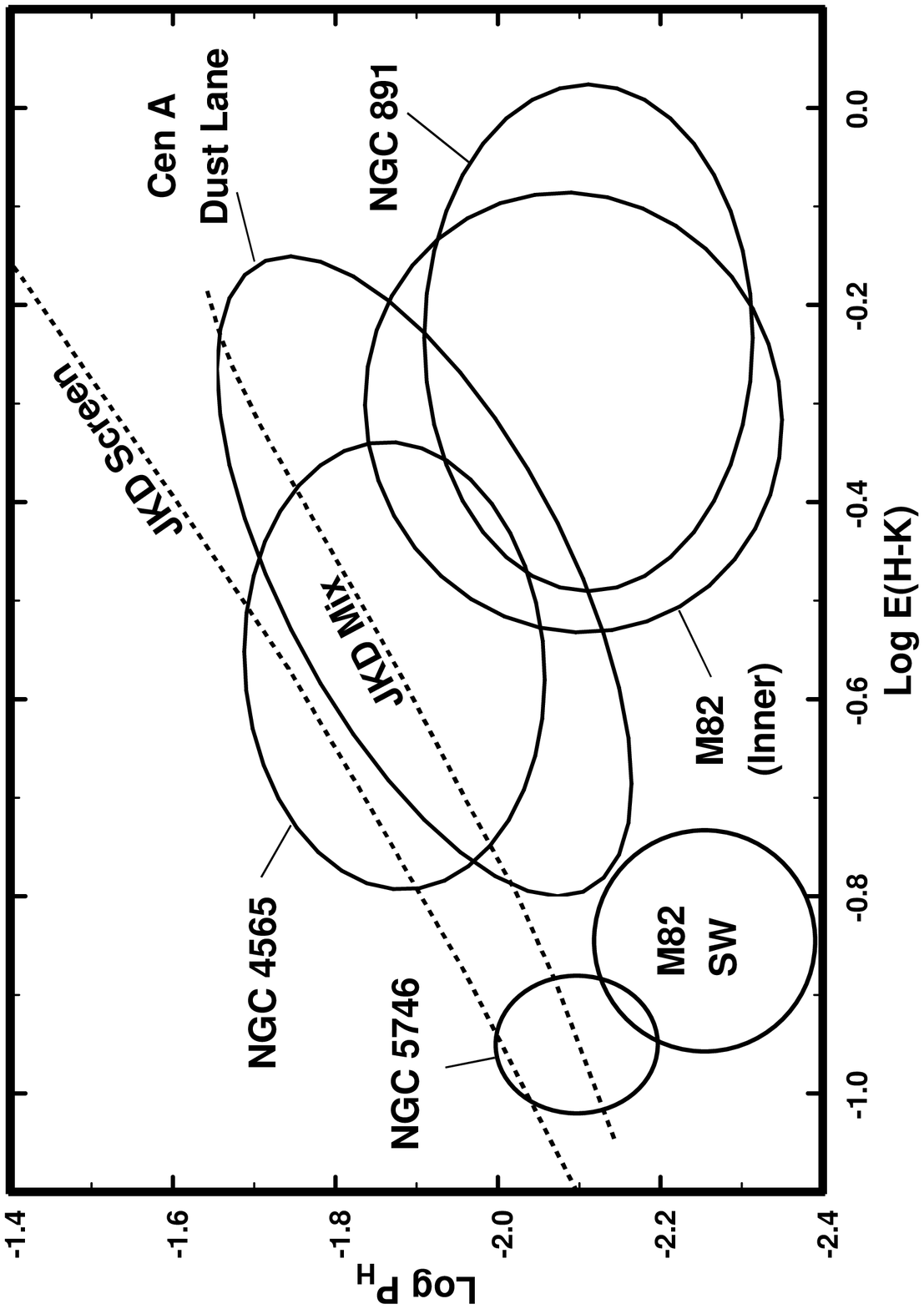]{A plot of polarzation strength vs. extinction
for several galaxies with imaging polarimetry at H and K. \label{fig9}}

\end{document}